\documentstyle[aps,epsfig]{revtex}
\newcommand{\be}{\begin{equation}}
\newcommand{\ee}{\end{equation}}
\begin{document}
\draft
\title{\bf A Proposed New Test of General Relativity and a Possible Solution
to the Cosmological Constant Problem}
\author{  Murat \"Ozer \footnote{E-mail: $m_{\_}h_{\_}$ozer@hotmail.com}} 
\address{CIENA Corporation, 991-A Corporate Boulevard\\
Linthicum MD 21090-2227\\}

\date{\today}
%
\maketitle
\begin{abstract}

\noindent Following a conjecture of Feynman, we explore the possibility
that only those energy forms that are associated with (massive or massless) 
particles  couple
to the gravitational field, but not others. We propose an
experiment to deflect electrons by a small charged sphere to determine if
the standard general relativity or this modified one 
corresponds to reality. The outcome of this experiment may also solve
the cosmological constant problem.
\end{abstract}\vspace{0.5cm}

The equivalence of mass and energy, expressed in his celebrated formula
$E=mc^2$, led Einstein to postulate that the energy-momentum tensor
$T^{\mu\nu}$ in the field equation of general relativity \cite{eins}
\be 
\label{eq:standard-gr}
R^{\mu\nu}-\frac{1}{2}g^{\mu\nu}R=\frac{8\pi G}{c^4}T^{\mu\nu},
\ee
contains all kinds of energies, such as matter, radiation, electromagnetic,
vacuum, etc.
Thus assuming that the vacuum energy is negligible, the field equation
(\ref{eq:standard-gr}) outside an object of total mass $M$ and static
electric charge $Q$ containg no neutral and charged masses or other fields
around it reduces to
\be 
\label{eq:reiss-nord}
R^{\mu\nu}=\frac{8\pi G}{c^4}T^{\mu\nu}_{EM},
\ee
where $T^{\mu\nu}_{EM}$ is the traceless energy-momentum tensor of the
electric field due to the charge $Q$ of the object. For the purpose of
this letter we shall classify different energy types into two. The first
class is the set of energy types with which massive or massless particles 
are associated. Thus the  energy of an already existing mass distribution
is obviously of this class
\footnote{The mass of the distribution may be constantly changing
due to its mechanical energy, its absorbtion or loss of heat energy, etc. 
Furthermore, while we can estimate how much electromagnetic binding energy
of an atom contributes to its rest mass we do not know, for example, how
much weak and gravitational energies contribute to it. We also know from
the E\"otv\"os experiment that electromagnetic binding energy contributes equally 
to inertial and gravitational massess. We assume this is the case for the
other energy types.  }
. Since the energy in an electromagnetic wave
(electromagnetic radiation) is carried in packages that behave like massless
particles (photons) the electromagnetic radiation energy is also of this 
class
\footnote{Recall that though massless, an `effective mass' can be assigned 
to photons.}
. Energies associated with other massless particles like neutrinos and
gravitons are further examples. Each energy type in this class may rightly 
be called `mass energy' or `particle energy'. The second class is the set of
energy types with which no particles are associated
\footnote{Of course, there is an `equivalent mass' through $E=mc^2$
for such energies too. But the crucial point is that there is no already
existing massive or massless particles associated with them. }
. The energies in the electric fields of a static charge distribution and 
between the plates of a capacitor as well as the
vacuum energy are of the second class.

There is plenty of emprical proof, such as the successes of the big-bang
cosmology and the deflection of light by the sun, that the first class
energies couple to the gravitational field. But there does not exist any
emprical proof at present for the coupling of the second class energies
to the gravitational field. Therefore, we do not know with certainty
if Eq. (\ref{eq:reiss-nord}) corresponds to a fact of nature. It lacks
experimental support. There is the intriguing possibility that we shall
consider in this letter, as first
hinted by Feynman \cite{feyn} when he said {\em '...Now gravity is supposed 
to interact with every
form of energy and should interact then with this vacuum energy. And 
therefore, so to speak, a vacuum would have a weight-an equivalent mass 
energy-and would produce a gravitational field. Well, it doesn't! The 
gravitational field produced by the energy in the electromagnetic field in a
vacuum-where there's no light, just quiet, nothing-should be enermous, so 
enermous, it would be obvious. 
The fact is, it's zero! Or so small that it's completely in disagreement 
with what we'd expect from the field theory. This problem is sometimes 
called the cosmological constant problem. It suggests that we're missing 
something in our formulation of the theory of gravity. It's even possible 
that the cause of the trouble-the infinities-arises from the gravity 
interacting with its own energy in a vacuum. And we started off wrong 
because we already know there's something wrong with the idea that gravity 
should interact with the energy of a vacuum. So I think the first thing we
should understand is how to formulate gravity so that it doesn't interact 
with the energy in a vacuum...'} According to this conjecture of Feynman  there 
is the possibility that the right side of Eq. (\ref{eq:reiss-nord}) 
may be zero:
\be 
\label{eq:new}
R^{\mu\nu}=0.
\ee
The importance of confronting with experiment the predictions of equations 
(\ref{eq:reiss-nord}) and (\ref{eq:new}) is not merely academic. If it turns
 out that Eq. (\ref{eq:new}) is the one favored by nature, we then have a 
very simple solution \cite{feyn} to the cosmological constant 
problem \cite{cosmo}. It would mean that being of the second class, the 
vacuum energy does not
couple to the gravitational field. The present value of the vacuum energy
density is as large as it had been in the early universe. The cosmological
constant, however, is simply zero, as it has always been. 

The purpose of this letter is to propose a deflection of electrons
by a positively charged sphere experiment to distinguish between equations
(\ref{eq:reiss-nord}) and (\ref{eq:new}). To this end, we shall need the
solutions of these equations. The solution of Eq. (\ref{eq:reiss-nord})
for a static and spherical distribution of mass $M$ and electric charge 
$Q$ located at $r=0$ is known as the Reissner-Nordstr{\o}m solution
\cite{reiss,nord}. It is given by
\begin{eqnarray} 
ds^2=\left(1-2\frac{GM}{c^2r}+\frac{Gk_eQ^2}{c^4r^2}\right)c^2dt^2-
\left(1-2\frac{GM}{c^2r}+\frac{Gk_eQ^2}{c^4r^2}\right)^{-1}dr^2-\nonumber\\
r^2d\theta^2-r^2sin^2\theta d\phi^2,
\label{eq:RN-solution}
\end{eqnarray}
where $k_e$ is the electric(Coulomb) constant. It should be noted that
according to Eq. (\ref{eq:new}), the electric field of the sphere does not
contribute to its gravitational field and hence must assert itself separately
and independently. Therefore, for weak fields Eq. (\ref{eq:new}) must
reduce to Laplace's equation
\be 
\label{eq:laplace}
\nabla^2(\Phi_G+\Phi_E)=0,
\ee
where $\Phi_G$ and $\Phi_E$ are the gravitational and electric potentials
of the sphere. Finding the solution of Eq. (\ref{eq:new}) proceeds along
the lines of the Schwarzschild solution \cite{sch}. We find
\begin{eqnarray} 
ds^2=\left(1-2\frac{GM}{c^2r}-2\frac{e}{m}\frac{k_eQ}{c^2r}\right)c^2dt^2
-\left(1-2\frac{GM}{c^2r}-2\frac{e}{m}\frac{k_eQ}{c^2r}\right)^{-1}dr^2
-\nonumber \\
r^2d\theta^2-r^2sin^2\theta d\phi^2,
\label{eq:newmetric}
\end{eqnarray}
where $-e$ and $m$ are the charge and the mass of an electron-a test
particle-in the viscinity of the spherical object
\footnote{Eq. (\ref{eq:newmetric}) can also be obtained intuitively by
classical energy considerations. Consider an electron moving radially away
from a sphere of mass $M$ and charge $Q$. For the electron to escape from
this object at a distance $r$ from its center and reach infinity with zero
speed, the escape velocity $v_{esc}$ satisfies
$mv^2_{esc}/2-GmM/r-ek_eQ/r=0.$ Replacing $v_{esc}$ with $c$, the speed of
light, (so that the electron cannot escape from the surface of radius r)
and dividing it by $mc^2/2$ the left side of this equation becomes
$(1-2GM/c^2r-2ek_eQ/mc^2r)$, which is the $g_{00}$ in Eq. (\ref{eq:newmetric})
.}
. To find the trajectory of an electron deflected by a positively charged 
sphere we also need the equations describing the trajectory according to
equations (\ref{eq:reiss-nord}) and (\ref{eq:new}). They are
\be 
\frac{d^2x^{\mu}}{ds^2}+\Gamma^{\mu}_{\alpha\beta}\frac{dx^{\alpha}}
{ds}\frac{dx^{\beta}}{ds}=-\frac{e}{mc^2}F^{\mu}_{\alpha}
\frac{dx^{\alpha}}{ds},
\label{eq:traject-RN}
\ee
according to Eq. (\ref{eq:reiss-nord}), and
\be 
\frac{d^2x^{\mu}}{ds^2}+\Gamma^{\mu}_{\alpha\beta}\frac{dx^{\alpha}}
{ds}\frac{dx^{\beta}}{ds}=0,
\label{eq:traject-new}
\ee
according to Eq. (\ref{eq:new})
\footnote{Note that charged particles follow the geodesics, Eq. (\ref
{eq:traject-new}), of the metric $g_{\mu\nu}$. This is a consequence of
Eq. (\ref{eq:new}). Note also that there is a different metric for
each particle with a different charge-to-mass ratio. The resulting 
theory, therefore, is a multi-metric theory.}. 
Here $\Gamma^{\mu}_{\alpha\beta}$ and
$F_{\mu\alpha}=\partial A_{\alpha}/\partial x^{\mu}-\partial A_{\mu}/
\partial x^{\alpha}$ are the connection coefficients and the electromagnetic
field strength tensor, with $A_{\mu}=(k_eQ/r, 0)$ being the electromagnetic
four-potential of the sphere. Before we indulge in obtaining the orbit 
equations in experimentally relevant form, we propose the following experiment. 
Consider a rectangular vacuum chamber. Let a small metallic
sphere of radius $R\approx 2-5\, cm$ positively charged to a voltage
$V(R)=k_eQ/R$ be hanged freely from an insulating thread. Let an electron gun
be located at a distance $d$ away from the equator (the $\theta=\pi/2$ plane)
of the sphere with an impact parameter $b$ which is the horizontal distance
between the initial path of the ejected electron beam and the center of the
sphere. Thus the initial position of the beam is $(x_i, y_i)=(-b, d)$ at an
angle $\phi_i=\pi/2+arctan(b/d)$. Put a calibrated fluorescent screen on the
negative $y$ axis at $\phi=3\pi/2$. The initial conditions for solving
the differential equations that we shall obtain (see equations 
 (\ref{eq:RNorbit-l}) and (\ref{eq:MGRorbit-l}) )
are $u(\phi_i)=r^{-1}_i$ and $du/d\phi(\phi_i)=
\sqrt{r^2_i/b^2-1/r_i}$ with $r_i=\sqrt{b^2+d^2}$. Make a large enough glass
window on the side of the box facing the screen (or monitor the position of 
the electron beam on the screen electronically). Compare the reading 
of the position of the beam with the predictions of the equations that we 
obtain now. 

Using spherical coordinates, we write the line element in the form
\be 
ds^2=e^{\eta}c^2dt^2-e^{-\eta}dr^2-r^2d\theta^2-r^2sin^2\theta d\phi^2.
\label{eq:ds-expo}
\ee
Inserting $d\theta/ds=0$ in Eq. (\ref{eq:traject-RN}) and integrating
the equations obtained for the coordinates $x^0=ct$ and $x^3=\phi$ we
get
\be 
\frac{dt}{ds}=\frac{e^{-\eta}}{c}\left(-\frac{qk_eQ}{mc^2}\frac{1}{r}
+a\right),
\label{eq:dtds-RN}
\ee
\be 
\label{eq:h}
r^2\frac{d\phi}{ds}=h,
\ee
where $a$ and $h$ are integration constants. Using equations 
(\ref{eq:dtds-RN}) 
and (\ref{eq:h}) in the equation obtained from the condition of timelike 
geodesics $g_{\mu\nu}(dx^{\mu}/ds)(dx^{\nu}/ds)=1$,
 putting $e^{\eta}\approx 1$
\footnote{For a sphere of $M=1kg$, $R=5cm$, $V(R)=10^3V$, we have for an
electron just grazing the sphere $g^{RN}_{00}=(1-2m_G/R+GV(R)^2/k_ec^4)=
(1-1.48\times 10^{-26}+9.19\times 10^{-49})\approx 1$.}
, and then differentiating with respect to
$du/d\phi$ we get
\be 
\frac{d^2u}{d\phi^2}+u=\frac{m_E}{h^2}+\frac{m_E^2}{h^2}u,
\label{eq:RNorbit-h}
\ee
where $u=1/r$ and the constant $a$ has been set to $1$ so that when $h=
l/mc$, with $l=mr^2\dot\phi$ being the ordinary angular momentum, the 
first term on the right side of Eq. (\ref{eq:RNorbit-h}) agrees with the 
corresponding Newtonian expression. Here $m_E=ek_eQ/mc^2=eRV(R)/mc^2$ has
the dimension of length and corresponds to $m_G=GM/c^2$ in the
Schwarzschild solution. On the other hand, we obtain from Eq. (\ref{eq:traject-new})
\be 
\label{eq:dtds-new}
\frac{dt}{ds}=\frac{e^{-\eta}}{c}
\ee
and Eq. (\ref{eq:h}) remains intact. By putting $e^{\eta}=(1-2m_G/r-2m_E/r)
\approx (1-2m_E/r)$ and proceeding as above we get
\be 
\frac{d^2u}{d\phi^2}+u=\frac{m_E}{h^2}+3m_Eu^2.
\label{eq:MGRorbit-h}
\ee
$mch$ is the conserved angular momentum of the electron in its rest 
frame. $h$ can be expressed in terms of $l$, the angular momentum in the 
laboratory frame, using equations (\ref{eq:dtds-RN}) and (\ref{eq:h}) in 
the Reissner-Nordstr{\o}m case and equations (\ref{eq:dtds-new}) and 
(\ref{eq:h}) in our case. They are, respectively, $h=l(1+m_Eu)/mc$ and 
$h=l(1-2m_Eu)^{-1}/mc$.
Inserting these in equations (\ref{eq:RNorbit-h}) and (\ref{eq:MGRorbit-h})
we finally obtain
\be 
\frac{d^2u}{d\phi^2}+u=\frac{m^2c^2}{l^2}\frac{m_E}{\left(1+m_Eu\right)},
\label{eq:RNorbit-l}
\ee
which is the orbit equation for the Reissner-Nordstr{\o}m solution, and
\be 
\frac{d^2u}{d\phi^2}+u=\frac{m^2c^2}{l^2}m_E\left(1-2m_Eu\right)^2+
3m_Eu^2,
\label{eq:MGRorbit-l}
\ee
which is the orbit equation in our case which we call modified general
relativity. By using the initial conditions 
stated above these equations can be solved numerically for $r=1/u$, the
predicted position of the electron beam on the screen from the center
of the sphere. By comparing the experimental value with these predictions,
the correct theory can be determined. In Figures 1 and 2 we depict the 
trajectory of the electrons according to the two theories. It is seen 
that the difference between the two predictions is large enough. Hence the
experimentally favored one can be picked up rather easily. 

Before we conclude, we wish to clarify the implications of the E\"otv\"os
experiment in regard to the theory presented here. The electromagnetic
energy of the atoms, or any other form of energy, in an object has
already been converted to mass ( and thus belongs to the first type
in our classification of the energy types above). What the E\"otv\"os
experiment tells us is that the electromagnetic energy contributes in
equal amounts to gravitational and inertial masses. It does not tell
us that this energy couples to the gravitational field independently
as energy. Had the electromagnetic 
energy coupled to the gravitational field independently,  a deflection
in the balance of the E\"otv\"os apparatus would have been seen
when two equal massess having 
considerably different electromagnetic binding energies were used. Of course
this does not happen. 

In conclusion, we have explored the conjecture of Feynman on a reformulation
of general relativity. In this new scheme only the `mass energy' couples
to the gravitational field, but not other energy forms. We have proposed a
deflection of electrons by a charged sphere experiment. The significance
of this experiment is that it not only provides a new test of general
relativity but also may point out to the solution of the cosmological
constant problem. Not being a wave, the energy of the vacuum is not 
associated with quantized packages and a formula like $E\propto f$, with
$f$ being the frequency, cannot be written and an effective mass 
$m_{eff}\propto f/c^2$ cannot be defined. Thus according to the scheme
presented here the field equation for vacuum is $R^{\mu\nu}=0$, implying
that the cosmological constant $\lambda=(8\pi G/c^4)\rho_V$ of standard
general relativity, with $\rho_V$ being the vacuum energy density, is
$\lambda=0\times\rho_V=0$ here. $\lambda$ has always been equal to zero!
Keeping on mind that (i) standard general relativity remains one of the least
tested of scientific theories, and (ii) the theory presented here offers
a very simple solution to the cosmological constant problem, the immediate
 performance of the experiment suggested here cannot be overemphasized.

We wish to thank Dr. Bahram Mashhoon for an e-mail correspondence on
the implications of the E\"otv\"os experiment.

%
\begin{figure}
\epsfig{figure=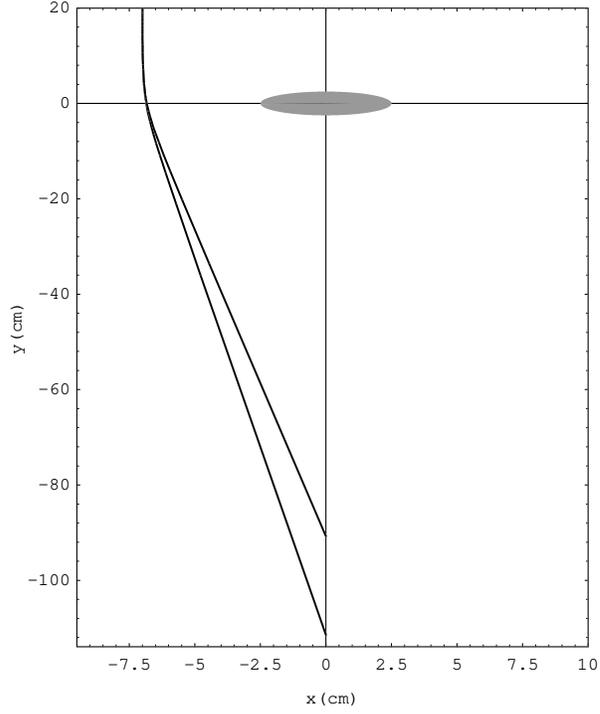, width=14cm}
\caption{The trajectories of the electron beam according to the
Reissner-Nordstr{\o}m (the bottom curve) and
the Modified General Relativity (the top curve) theories for an
anode-cathode voltage of $30\, kV$ for the electron gun located at a vertical
distance of $20\, cm$ with an impact parameter of $7\, cm$, for a sphere of
$R=2.5\, cm$ and $V(R)=5\, kV$.
\label{fig:Figure 1} }
\end{figure}
\begin{figure}
\epsfig{figure=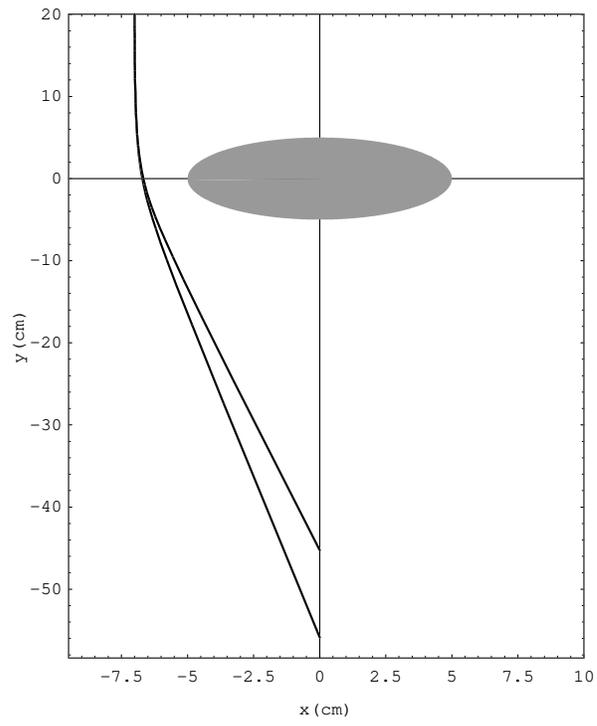, width=14cm}
\caption{Same as Fig.1, but $R=5\, cm$.
\label{fig:Figure 2} }
\end{figure}

\end{document}